\documentclass[twocolumn,trackchanges]{aastex63} %linenumbers

\usepackage{subfigure}
\usepackage{CJK}

\graphicspath{{./}{figures/}}

\shorttitle{The north-south Flaring}
\shortauthors{Yu, Wang \& Cui et al.}

\begin{document}

\begin{CJK*}{UTF8}{gbsn}
\title{The Flare and Warp of the Young Stellar Disk  traced with LAMOST DR5 OB-type stars}
\author{Yang Yu(于扬)}
\affil{Equal first authors}
\affil{Department of Physics, Hebei Normal University, Shijiazhuang 050024, China}

\author[0000-0001-8459-1036]{Hai-Feng Wang(王海峰)}
\affil{Equal first authors}
\affil{GEPI, Observatoire de Paris, Universit\'e PSL, CNRS, Place Jules Janssen 92195, Meudon, France}
\affil{South-Western Institute for Astronomy Research, Yunnan University, Kunming, 650500, China}
\affil{National Astronomical Observatories, Chinese Academy of Sciences, Beijing 100101, China}

\author[0000-0003-1359-9908]{Wen-Yuan Cui(崔文元)}
\affil{Department of Physics, Hebei Normal University, Shijiazhuang 050024, China}

\author{Lin-Lin Li(李林林)}
\affil{Department of Physics, Hebei Normal University, Shijiazhuang 050024, China}

\author[0000-0002-1802-6917]{Chao Liu(刘超)}
\affil{National Astronomical Observatories, Chinese Academy of Sciences, Beijing 100101, China}

\author[0000-0002-6434-7201]{Bo Zhang(章博)}
\affil{National Astronomical Observatories, Chinese Academy of Sciences, Beijing 100101, China}
\affil{Department of Astronomy, Beijing Normal University, Beijing, 100875, China}

\author[0000-0003-3347-7596]{Hao Tian(田浩)}
\affil{National Astronomical Observatories, Chinese Academy of Sciences, Beijing 100101, China}

\author{Zhen-Yan Huo(霍振燕)}
\affil{Department of Physics, Hebei Normal University, Shijiazhuang 050024, China}

\author{Jie Ju(巨洁)}
\affil{Department of Physics, Hebei Normal University, Shijiazhuang 050024, China}

\author{Zhi-Cun Liu(柳志存)}
\affil{National Astronomical Observatories, Chinese Academy of Sciences, Beijing 100101, China}

\author{Fang Wen(温芳)}
\affil{Department of Physics, Hebei Normal University, Shijiazhuang 050024, China}

\author{Shuai Feng(冯帅)}
\affil{Department of Physics, Hebei Normal University, Shijiazhuang 050024, China}

\correspondingauthor{Wen-Yuan Cui(崔文元)}
\email{wenyuancui@126.com, cuiwenyuan@hebtu.edu.cn}

\begin{abstract}

We present analysis of the spatial density structure for the outer disk from 8$-$14 \,kpc with the LAMOST DR5 13534 OB-type stars and {\bf observe similar flaring on north and south sides of the disk implying that the flaring structure is symmetrical about the Galactic plane, for which the scale height at different Galactocentric distance is from 0.14 to 0.5 \,kpc}. By using the average slope to characterize the flaring strength we find that the thickness of the OB stellar disk is similar but flaring is slightly stronger compared to the thin disk as traced by red giant branch stars, possibly implying that secular evolution is not the main contributor to the flaring but perturbation scenarios such as interactions with passing dwarf galaxies should be more possible.  When comparing the {\bf scale height of} OB stellar disk of the north and south sides with the gas disk, the former one is slightly thicker than the later one by $\approx$ 33 and 9 \,pc, meaning that one could tentatively use young OB-type stars to trace the gas properties. Meanwhile, we unravel that the radial scale length of the young OB stellar disk is 1.17 $\pm$ 0.05 \,kpc, which is shorter than that of the gas disk, confirming that the gas disk is more extended than stellar disk. What is more, by considering the mid-plane displacements ($Z_{0}$) in our density model we find that almost all of $Z_{0}$ are within 100 \,pc with the increasing trend as Galactocentric distance increases.

\end{abstract}


\section{Introduction} 

Disk galaxy structure and origins are central problems in galaxy formation and evolution. The star counts are commonly used method to explore the structure of the disk galaxy such as our home galaxy, Milky Way (MW), and it also has been proved as a very effective tool in revealing the structure and substructure such as flare, warp, truncation, spiral arms, etc, thus then it will push our understanding of the Galaxy formation and evolution \citep{vanderkruit2011, blandhawthorn2016}. A dual exponential model is widely used to study the spatial structure of disk: one for thin disk and the other one for thick disk \citep{gilmore1983, juric2008, bovy2012c, Rix2013, bovy2016, blandhawthorn2016}. An alternative form, that is, $sech$ function is also often used in the vertical density profile model \citep{vanderkruit1988,vanderkruit2011, wang2018b}. 

With the help of Red Giant Branch (RGB) stars, \citet{wang2018b} {\bf revealed} that the Galactic disk radial profile is composed by three sections of exponential law with  the scale length of $2.12\pm0.26$, $1.18\pm0.08$, and $2.72$\,kpc respectively at  $R<11$, $11\leq R\leq14$, and $R>14$\,kpc. Meanwhile, they also found the scale length of the thick disk is larger than that of the thin disk and some density asymmetries might be explained, to some extent, by shifting either the mid-plane displacements of the thin or the thick disk. What's more, they also detected clear flaring features, i.e., the scale height increases with the distance, from 8 to 19 \,kpc.

\citet{juric2008} {\bf found} that the scale length of the thin disk is shorter than that of the thick disk based on {\bf Sloan Digital Sky Survey}. Similar results were shown in \citet{Chen2017} by using Photometric data from the Xuyi Schmidt Telescope Photometric Survey of the Galactic Anti-Centre and the Sloan Digital Sky Survey. However, \citet{wan2017} suggested that the younger geometrical thin disk with 4.7 \,kpc of the exponential scale length is larger than the older thick disk with 3.4 \,kpc of the exponential scale length with two components, which has been argued by \citet{bovy2012c} using mono-abundance population methods. {\bf The geometry definition mentioned in this work is based on the density profile\citep{Martig2016}.}

The scale height of the thin disk is around 220--450 \,pc and the thick disk is between 700 and 1200 \,pc in the solar vicinity which are either from exponential model or $sech$ function for the vertical component~\citep{blandhawthorn2016}. As we move forward to the larger Galactic distance, the value increases due to the flaring effects, in another words, because the disk flares, the exponential scale height increases with increasing Galactocentric radius. Recently, with 250,000 OB stars containing all types of Gaia \citep{Gaia2016} and 2MASS \citep{Skrutskie2006} photometric catalog, \citet{Lichengdong2019} showed that the scale length is $2.10 \pm 0.01$ \,kpc and scale height is from 132$-$450 \,pc.

{\bf It appears that secular heating of the stellar disk caused by the scattering of the spiral structures~\citep{Sellwood1984, Carlberg1985} or giant molecular clouds~\citep{Jenkins1990, Jenkins1992} might cause the feature that the scale height of the younger populations is shorter than that of the older ones \citep{yu2017, liu2012}.}The well known flare was explained by migration of stars \citep{Solway2012}, the larger vertical excursions due to the decreased gravitational pull in the outer disk can lead to the flaring \citep{bovy2016}. An alternating explanation was the dynamical heating \citep{Quinn1993}. However, as first proposed by \citet{Minchev2012}, flaring by radial migration does happen, but the corresponding features are completely wiped out by mergers shown in Fig. 2 of \citet{Minchev2014}. Followed by this work, \citet{Minchev2015, Minchev2016, Minchev2018} suggested that migration suppresses flaring in the presence of external perturbations in CDM cosmology \citep{Quinn2009}. This is in agreement with some other cosmological simulations, e.g., \citet{Grand2016, Ma2017}. 

As far as we know about the scenarios for the flaring in general, one is the secular evolution, e.g., mentioned in the works of \citet{Minchev2012, Minchev2014, Narayan2002}, the other is perturbation, e.g., passing dwarf galaxies or satellites denoted in \citet{Kazantzidis2008, Villalobos2008, Laporte2018}, and so on.

It is almost a consensus that warp exists in the Milky Way, that is describing the disk bends upwards and downwards in the north and south hemisphere separately. The kinematical and spatial signals of the stellar warp along with radius and varies with the azimuth angle are shown in some works \citep{Chen2019, Liu173, Lop141}. The dynamical signals of the warp corresponding to the increasing trend of vertical velocity with vertical angular momentum was displayed in \citet{Schonrich2017, Huang2018}, and \citet{wang2020a} also proposed that most likely the main S-shaped structure of the warp is a long-lived nonsteady feature.

According to our current understanding, it seems that previous studies focused on the structure of disk tended to use older tracers for the reason that the number of old stars is quite large, especially for the outer disk, which makes it easier to describe the density distribution of the disk , e.g., \citet{bovy2016, lopez2002, lopez2014, wang2018b} and reference therein. Up to now we are still not sure the exact flaring mechanisms and the detailed flaring features in the north and south stellar disk. Before LAMOST (Large Sky Area Multi-Object fiber Spectroscopic Telescope; also known as Guo Shou Jing telescope) and Gaia, we could not get large sample for the OB stars due to some reasons such like sample purity, sampling rates of disk region, precise distance, atmospheric parameters, line indices, high precision photometry and so on. In this work, we use young OB-type stars to explore spatial structure of the young thin disk, which will help us to better understand the formation of the stellar disk. 

The structure of this paper is as follows. In Section 2, we describe how we select OB-type stars and correct the selection effects. In Section 3, we introduce the vertical star counts model for this work. In Section 4, we show the results about the flaring disk and discussions. Finally, we summarize this work in Section 5.

\section{sample and selection effects}  

\subsection{The OB-type stars catalog }

LAMOST sky survey \citep{cui2012, Deng2012, zhao2012} has observed a total of 4,154 sky regions and released 9,026,365 spectra for the phase-I. The DR5 catalog includes spectral parameters of 5,348,712 stars with Teff, logg, and [Fe/H]. Now LAMOST is being conducted the second 5$-$year survey since 2018 with the Medium Resolution of 7,500.

OB-type stars is composed of massive ($M > 8M_{\odot}$) stars, hot sub-dwarfs, white dwarfs, and post-AGB stars, etc. Among them, the massive stars have not moved out of the star-forming region because of their short life. \citet{Liu2019} identified 16,032 OB-type stars selected by spectral line indices space, including 22,901 spectra with signal-to-noise ratio larger than 15 in the g band, 948 hot subdwarf spectra, and 160 white dwarf spectra. The sample in this catalog has been tested by independent human eyes one by one and thus has high accuracy. The completeness for OB stars earlier than B7 contained in LAMOST DR5 is better than 89 ± 22\%, which is used here, it is by now the largest spectroscopic OB-type star sample. This sample has been used in \citet{Cheng2019, wang2020d} for disk asymmetries. 

\begin{figure}[htbp]
\centering
\subfigure{ \includegraphics[width=0.5\textwidth]{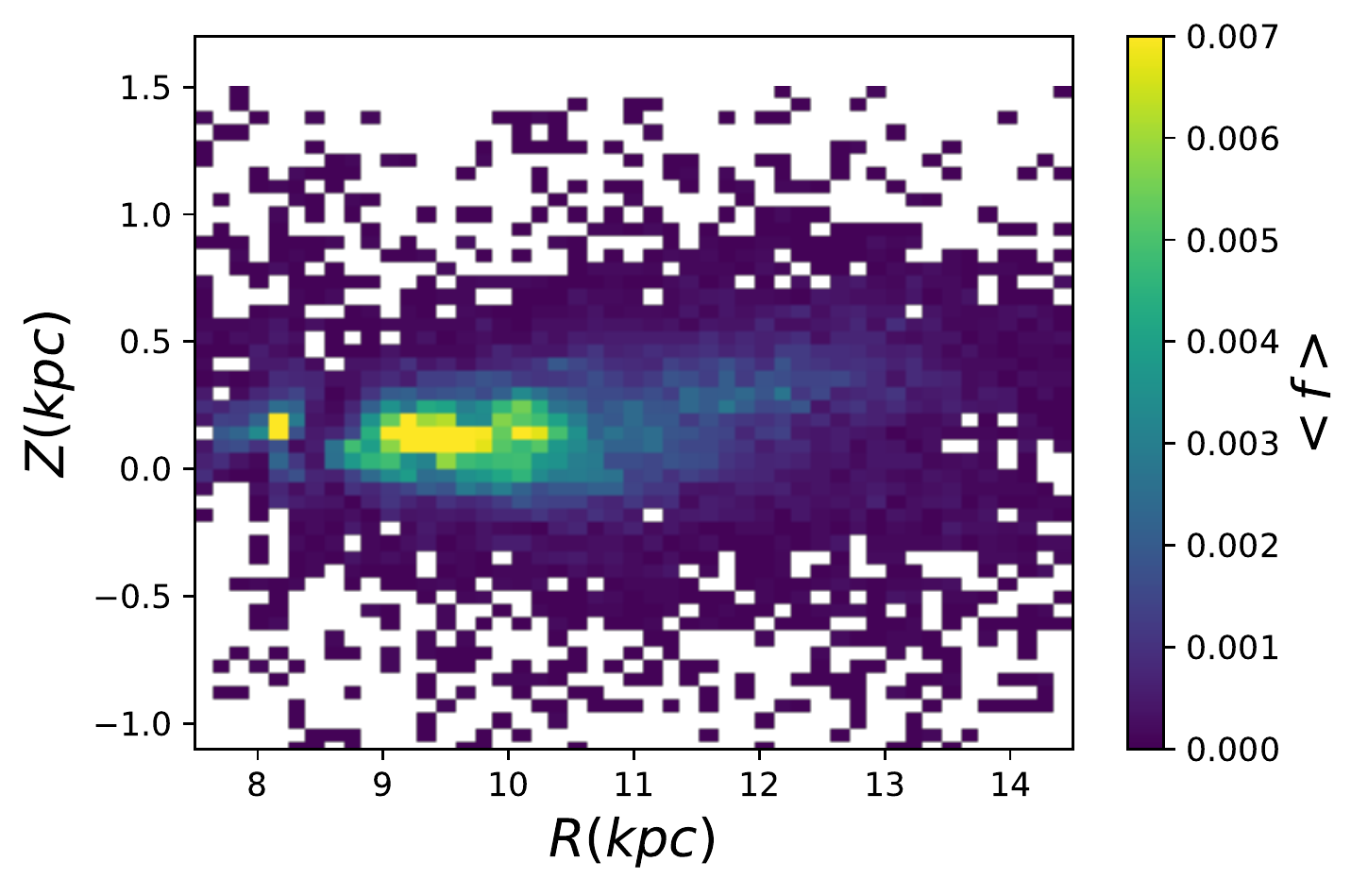}}
\quad
\subfigure{\includegraphics[width=0.4\textwidth]{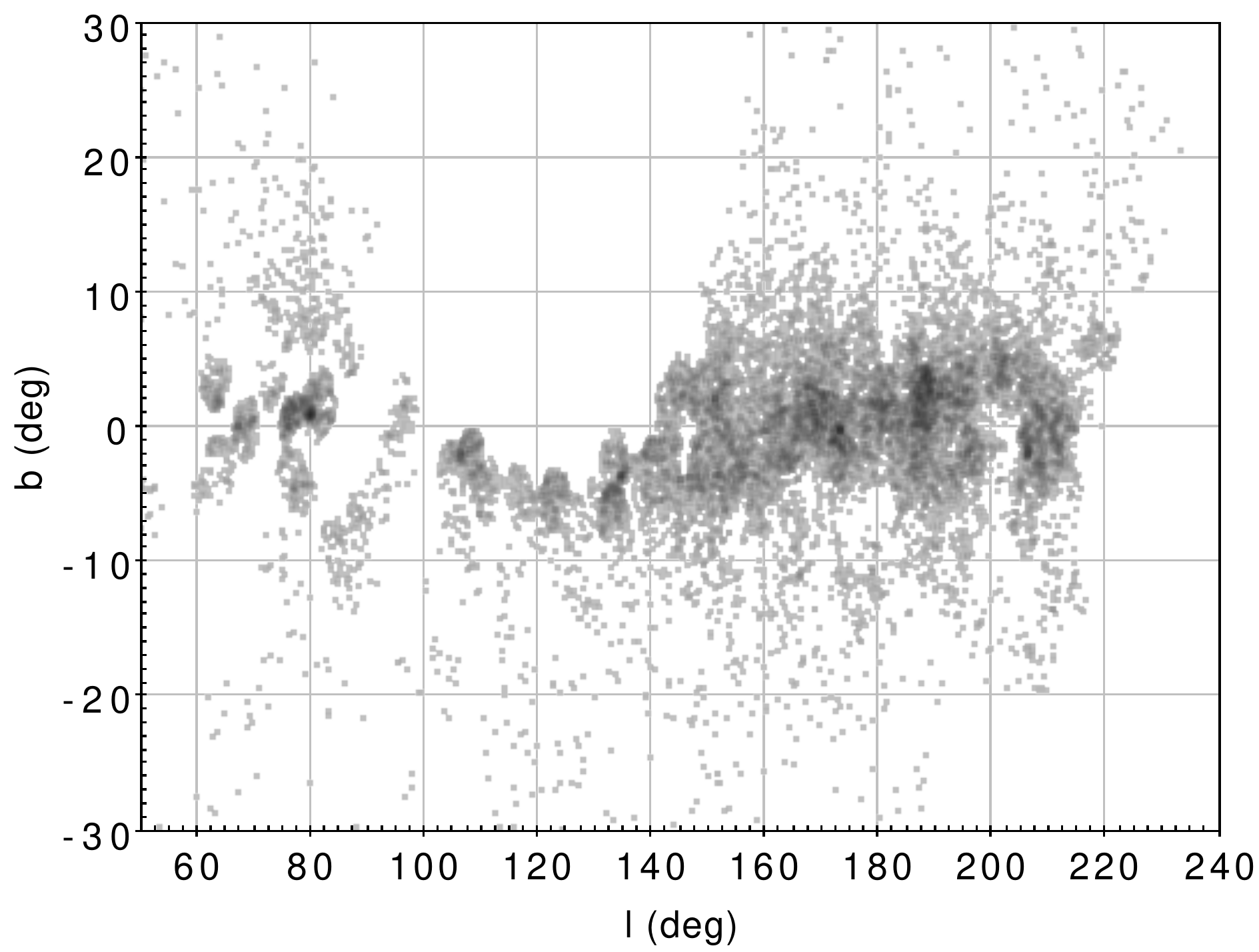}}
\caption{The top one of the figure shows normalized number counts distribution of OB type stars without considering selection effects during this work.The bottom one is the  sky coverage of our OB-type stars adopted from \citet{Liu2019} and the area of the sky mainly comes from the Galactic-Anti Center direction, the sample is mainly from $l$ = [140 210] deg, $b$ = [-3 5] deg with the relatively small contribution from other areas}.
\label{counts}
\end{figure}

We cross match the OB-type star catalog with Gaia DR2 \citep{Gaia2018a} so that the distance of around 16,000 stars can be obtained. We suggest the small parallax zero$-$point bias will not affect our final conclusion due to that, our sample are actually mainly within 5 \,kpc away from the sun and the parallax is larger than 0.2 \,mas, the small {\bf parallax}  zero$-$point bias around 0.05 \,mas or even small can not change our conclusion, {\bf the zero point bias increases with distance for Gaia, it is very small within 4-5 \,kpc \citep{Gaia2018b}, that is our range}. Even though, we still have already finished a test for the results to prove our conclusions are robust. The sky coverage and the number counts distributions of the sample we used in this work are denoted in Fig.~\ref{counts}, respectively. 

\subsection{The OB-type stars selection correction}

We adopt the method to correct the selection effect developed by \citet{liu2017a} {\bf and it was tested by simulation data}, this method has been used in several works based on LAMOST giant stars, e.g., \citet{wang2018b, xu2017}. Before applying the selection correction, we further clarify the OB samples by adopting the below criteria:

1) Parallax/Parallax$_{error}$  $>$ 2.5 and parallax$>$0; 

2) $J$$_{error}$, $K$$_{error}$ $<$ 0.1\,mag; 

3) $K$ $<$ 14.3 \,mag; 

After applying these criteria in consideration of the parameter accuracy and the limited magnitude or the completeness of the photometric survey, there are 13,534 OB-type stars left during this work. Then we assume that, at a given line-of-sight of Galactic coordinates ($l$, $b$) and color magnitude box, the probability finding a star at distance $D$ in the LAMOST survey data should be roughly same as in the photometric  {\bf 2MASS \citep{Skrutskie2006}} data. Based on these assumptions and Bayesian theory we could obtain a selection factor, thus then we could calculate the corrected density. Not only this method could be used in the spatial density, but also in mono-abundance population structures. Below we briefly introduce main principle to show how we correct selection effects:

\begin{equation}\label{eq:pp}
p_{ph}(D|J-K,K,l,b)=p_{sp}(D|J-K,K,l,b),
\end{equation}
where $J-K$ and $K$ stand for the colour index and $K_s$-band magnitude in {\bf 2MASS \citep{Skrutskie2006}}, respectively.

Then the stellar densities for photometric data, $\nu_{ph}$, and for the spectroscopic data, $\nu_{sp}$, are associated with each other through

\begin{eqnarray}\label{eq:nuPDF}
&\nu_{sp}(D|J-K,K,l,b)=\nonumber\\
&\nu_{ph}(D|J-K,K,l,b)S(J-K,K,l,b),
\end{eqnarray}
where $S$ represents for the selection function of the spectroscopic data. The selection function can be determined by
\begin{equation}\label{eq:S2}
S(J-K,K,l,b)={n_{sp}(J-K,K,l,b)\over{n_{ph}(J-K,K,l,b)}},
\end{equation}
where $n_{sp}$ and $n_{ph}$ are the star counts of the spectroscopic and photometric data in the $J-K$ vs. $K$ plane from 2MASS \citep{Skrutskie2006}, respectively. Then we could obtain the corrected stellar density by combining the selection factor and spectroscopic number density in the catalog.

Because the spectroscopic stars are usually very limited in a line-of-sight and the uncertainty of distance to the stars cannot be ignored, a kernel density estimation (KDE) is applied to derive $\nu_{sp}$ along a line-of-sight, i.e.

\begin{equation}\label{eq:KDE2}
\nu_{sp}(D|J-K,K,l,b)={1\over{\Omega D^2}}\sum_{i}^{n_{sp}(J-K,K,l,b)}{p_i(D)},
\end{equation}
where $p_i(D)$ is the probability density function of $D$ for the $i$th star.

{\bf In general, D is the distance from the sun and we normalize the probability for D along line of sight from 0 to infinity, the uncertainty of this method is about 25\% including extinction contribution, the size of each bin for star counts is $\triangle$ (J$-$K) = 0.1 and $\triangle$ K = 0.25, $J-K$ vs. $K$ plane is from 2MASS. Please see more details about this method in our previous papers \citep{liu2017a, wan2017, wang2018b, Liu2018}.}

\section{Star counts model} \label{sec:tables}

In this section, we will introduce how we construct the density model of the disk used in this work. The Galactic disk mainly consists of  two components of thin and thick disks \citep{gilmore1983}. However, the OB-type stars in our sample are all young stars with lower galactic latitude. Therefore, they almost could not contribute to the old thick disk so that we choose to only use a single exponential vertical profile. \citet{wang2018b} used the non-parametric method normalizing the radial density part into the mid-plane density, thus then they could simplify the 2D density plane to 1D density slice, to unravel the outer disk is more complex than what we thought before. During this work, we also use similar method but we consider the north and south sides separately and introduce the mid-plane displacement $Z_{0}$ as an additional free parameter, then we slice the OB-type star sample into bins along the direction of the galactic centric distance R, which are centered at $R$ $=$ 8.5, 9.5, 10.5, 11.5, 12.5, and 13.5 \,kpc. The width of each bin is 1 \,kpc by considering the sampling rates and poisson noise. 

We use normalized exponential model to fit the stellar density distribution assuming north-south vertical symmetry as follows:

\begin{equation}\label{numbermodel}
\nu(Z|R)=\left\{
\begin{array}{rcl}
\nu_0(R)e^{-\left(\frac{Z-Z_0}{h_n(R)} \right)} & & {Z > Z_0}\\
\nu_0(R)e^{-\left(\frac{|Z-Z_0|}{h_s(R)}\right)} & & {Z < Z_0}
\end{array} \right.
\end{equation}

In which, $\nu_{0}(R)$ is the total volume density when $Z = Z_{0}$,  $h_{n}(R)$ is the scale height on the north side of the young disk, $h_{s}(R)$ is the scale height on the south side of the disk and $Z_{0}(R)$ is the true mid-plane displacement. We assume $Z = Z_{0}$ is a free parameter in the data and the sample is separated by $b=0$ or $Z=0$ due to that we don't know the true value of the mid-plane displacement and we want to explore it using our method, then we could explore scale height of the north and south and true mid$-$plane $Z_{0}$ simultaneously. And up to now we are also not sure $Z = Z_{0}$ value at different distance and and it is not clear that whether the scale height of the north and south is similar or not. In order to derive all the unknown parameters at each $R$ bin, we firstly set up the histogram of the mean vertical stellar density along $Z$ grid. Then we sample the likelihood distribution as:

\begin{eqnarray}\label{lnlike}
  &\mathcal{L}(\{\nu_{\rm obs}(Z_i|R)\}|\nu_0, h_{n}, h_{s}, Z_{0}) =\nonumber\\ &\prod_i\exp\left[-{1\over{2}}(\nu_{\rm obs}(Z_i|R)-\right.\nonumber\\
  &\left.\nu_{\rm model}(Z_i|R, \nu_0, h_{n}, h_{s}, Z_{0}))^2\right],
\end{eqnarray}

Where $Z_{i}$ is the $i-$th point of the $Z$ grid, $\nu_{obs}$ is the stellar density calculated from the observation, $\nu_{model}$ is the stellar density of the theoretical model, and corresponding errors are calculated by bootstrap method. Here we have four free parameters in the fitting process: $ ln( \nu_{0}(R))$, $h_{n}(R)$, $h_{s}(R)$, $Z_{0}(R)$. 

\section{Results and discussions} \label{sec:autonumber}
\begin{figure}[htbp]
\centering
\subfigure{
\includegraphics[scale=0.35]{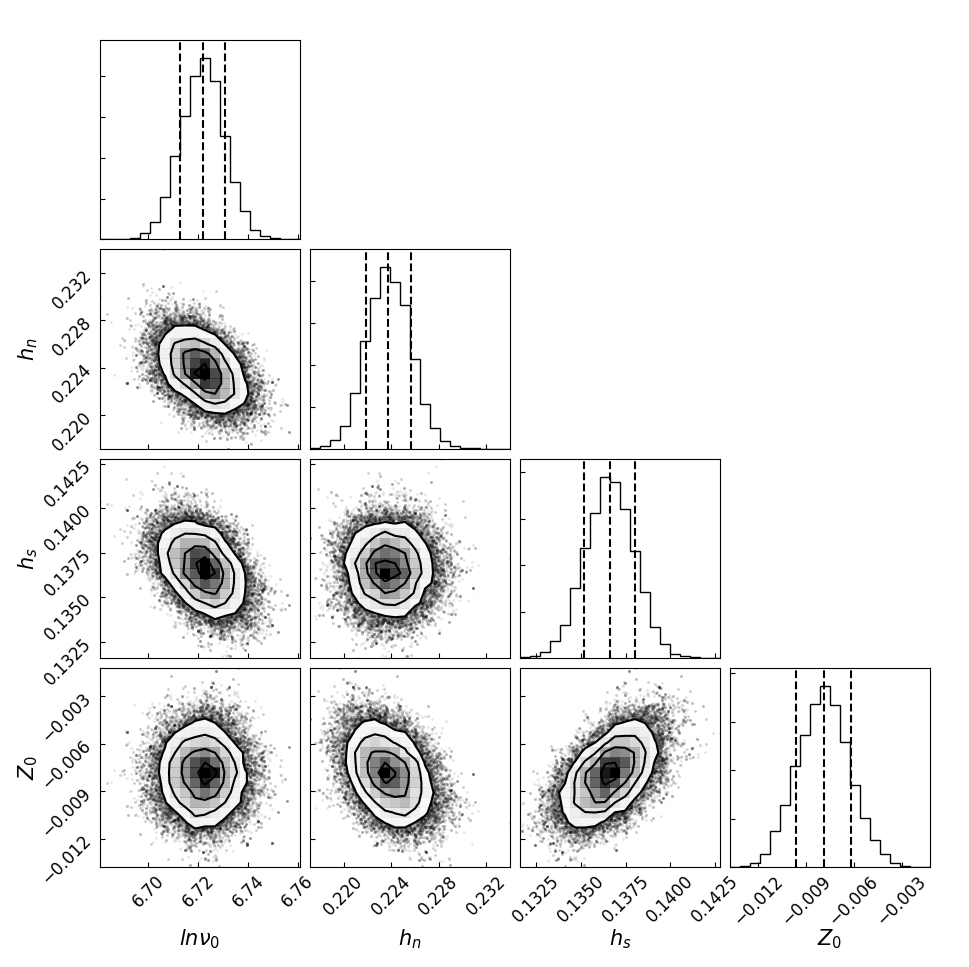} \label{1}}
 \quad
 \subfigure{\includegraphics[scale=0.13]{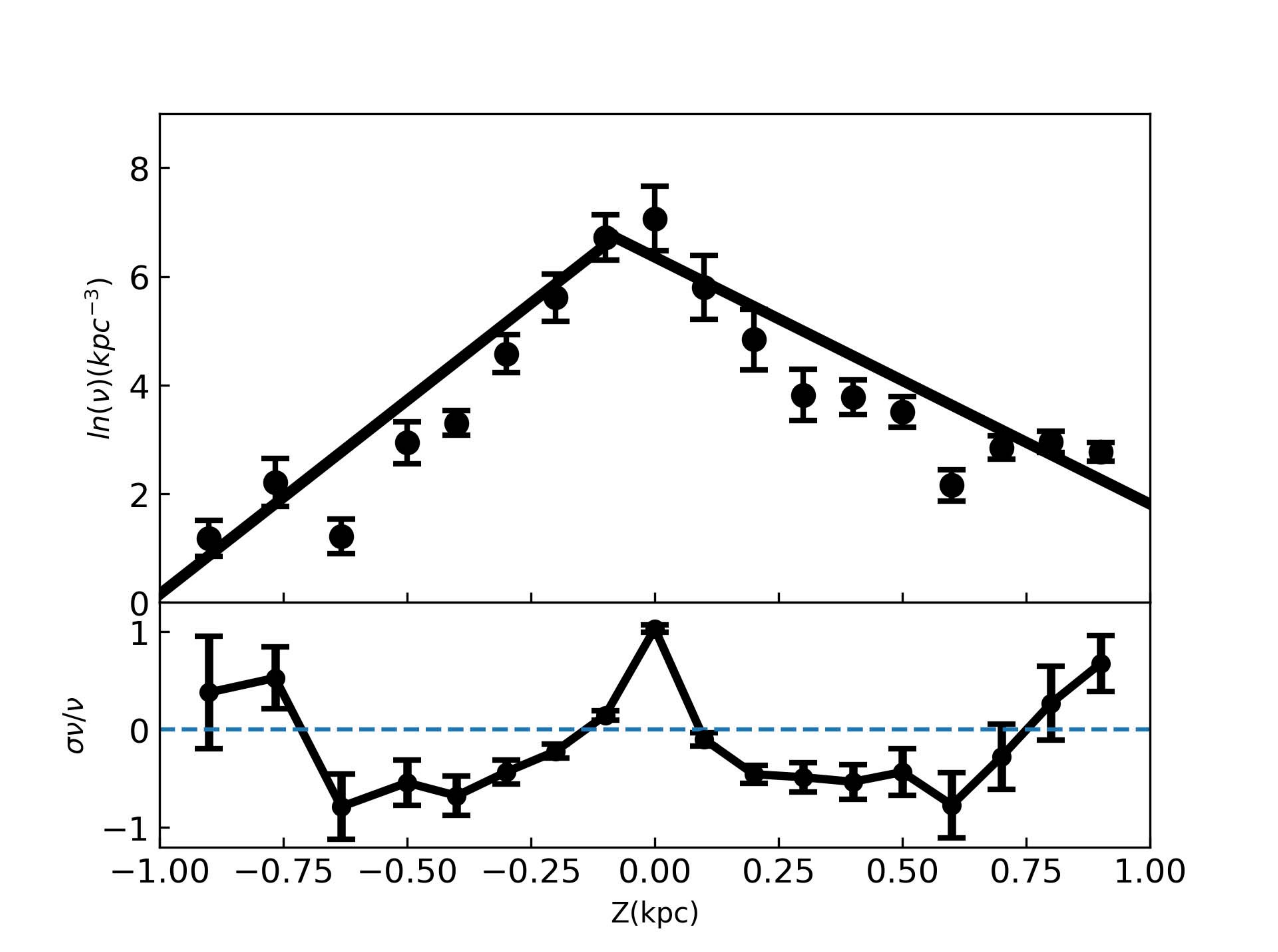} \label{2} }
\caption{The top panel shows the likelihood distribution of the parameters ($ln (\nu_{0})$ ($kpc^{-3}$),  $h_{z}$ (\,kpc), $Z_{0}$ (\,kpc)) for 8$-$9 \,kpc drawn from the MCMC simulation, the uncertainties in the estimates are determined using the 15th and 85th percentiles of the MCMC samples. The bottom one is the fitting results and relative errors of model and data, it appears that there are some oscillations in the solar neighborhood.}
\label{mcmc1}
\end{figure}

The Markov Chain Monte Carlo (MCMC) method provided by $emcee$ \citep{emcee2013} is used for sampling the likelihood distribution. The best fitting value of the free parameters is selected as the peak of the likelihood distribution and the estimated uncertainty is determined by the values of 15 and 85 percentiles of the likelihood distribution. The MCMC fitting results are shown in table 1 listing the four best fitting parameters at different $R$ bins. As an example, we show the likelihood distribution of the parameters for 8$-$9 \,kpc and the fitting results of model and data in  Fig.~\ref{mcmc1}.

\begin{deluxetable*}{ccccc}
\tablenum{1}
\tablecaption{Best fitting parameters of different $R$ bin with vertical density distribution models.\label{tab:messier}}
\tablewidth{0pt}
\tablehead{
\colhead{R(kpc)} & \colhead{$ln\nu_{0}(kpc^{-3})$} & \colhead{$h_{n}(kpc)$} & \colhead{$h_{s}(kpc)$} & \colhead{$Z_{0}(kpc)$}}
\decimalcolnumbers
\startdata
$8<R<9$ &6.72±0.009& 0.22±0.002& 0.14±0.001& -0.08±0.002 \\
$9<R<10$ & 6.011±0.005& 0.20±0.001& 0.15±0.001 &0.01±0.001 \\
$10<R<11$ &5.03±0.009& 0.17±0.002 &0.17±0.002& 0.02±0.001 \\
$11<R<12$ & 3.87±0.018& 0.23±0.003 &0.20±0.005& 0.03±0.004 \\
$12<R<13$ & 3.17±0.040&0.30±0.011& 0.30±0.023& 0.08±0.013 \\
$13<R<14$ & 2.24±0.100& 0.45±0.062& 0.50±0.126 &0.09±0.045  \\
%$14<R<15$ & 1.89±0.170& 0.78±0.176& 0.35±0.260& -0.14±0.086 \\
\enddata
\end{deluxetable*}

\subsection{The North-South flaring disk compared with stellar disk}\label{subsec:figures}

\begin{figure}
  \centering
  \includegraphics[width=0.5\textwidth]{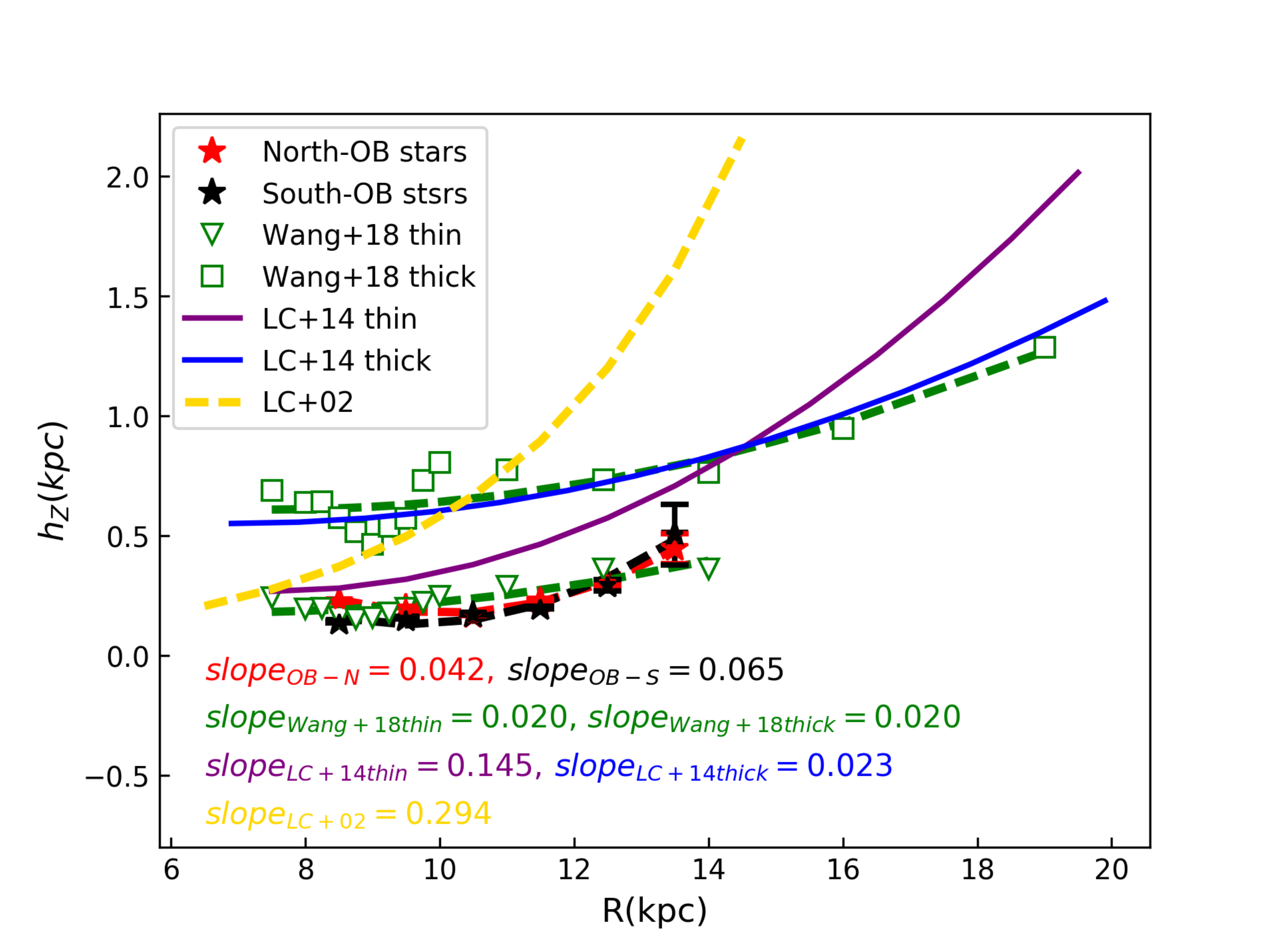}
  \caption{The figure shows the OB-type stars and other tracers scale height distribution along the radial distance, red pentagram line is northern OB-type stars scale height, black pentagram line is the southern OB-type stars scale height. Green triangle line is the thin disk scale height of \citet{wang2018b} and green rectangle line is the thick disk scale height of \citet{wang2018b}. Brown and blue line are the thin disk and thick disk of \citet{lopez2014} respectively. Yellow line  is the results of \citet{lopez2002}. These data compared here are originated from the same regions and the {\bf average slope describing the strength of flaring is labeled in the bottom left, different colors are corresponding to different results.}}
  \label{flaringns1}
\end{figure}

The scale height of the young stellar disk is shown in Fig.~\ref{flaringns1}. As we could see, the north and south scale height of the young stellar disk is increasing {\bf from 0.14 to 0.5 \,kpc in the range of 8$-$14 \,kpc}, that is the clear flaring signatures, {\bf so we could say we detect the features of the flaring in the young stellar disk.}Meanwhile, it is shown that the pattern of the north and south disk is similar or slightly different, the average difference of the scale height in the north and south is 19 \,pc and the slope of both sides has no large difference, suggesting us that the flaring of the disk might be symmetrical. {\bf During this work we use average slope of the scale height vs. distance to describe the flaring strength, which has been used in \citet{wan2017}.}

Previous works are also compared in the figure and we could see, the thickness of the young stellar disk is similar to or slightly different from the thin disk traced by RGB stars in \citet{wang2018b}. The clear difference between the young stellar disk of OB stars and the thin disk or disk distribution of \citet{lopez2002, lopez2014} are also displayed here. The reason for the difference could be naturally understood that the different populations and methods are used in these works, e.g., here we use the young OB-type stars and non-parametric method, \citet{lopez2002} used the photometric red clump stars without considering the two components of the disk, and \citet{lopez2014} adopted the F8V-G5V stars to study the changes of the scale height of thin and thick disks with different methods. Importantly, we detect the almost symmetrical flaring signature in the young stellar disk of OB samples, and {\bf it is similar with or even slightly stronger than the old stellar thin disk traced by RGB samples for slope value}. Please also notice that our OB star sample is distributed in the Galactic-Anti Center and we just use the projection distance, that is to say, $R$ and $Z$, and these compared data in Fig.~\ref{flaringns1} are almost originated from the same regions, mainly from the north hemisphere. There are some clear different regions for gas disk but we could still compare the disk scale length and scale height.

\subsection{The North-South flaring disk $h_{z}$ compared with gas disk}\label{subsec:figures}

\begin{figure}
  \centering
  \includegraphics[width=0.5\textwidth]{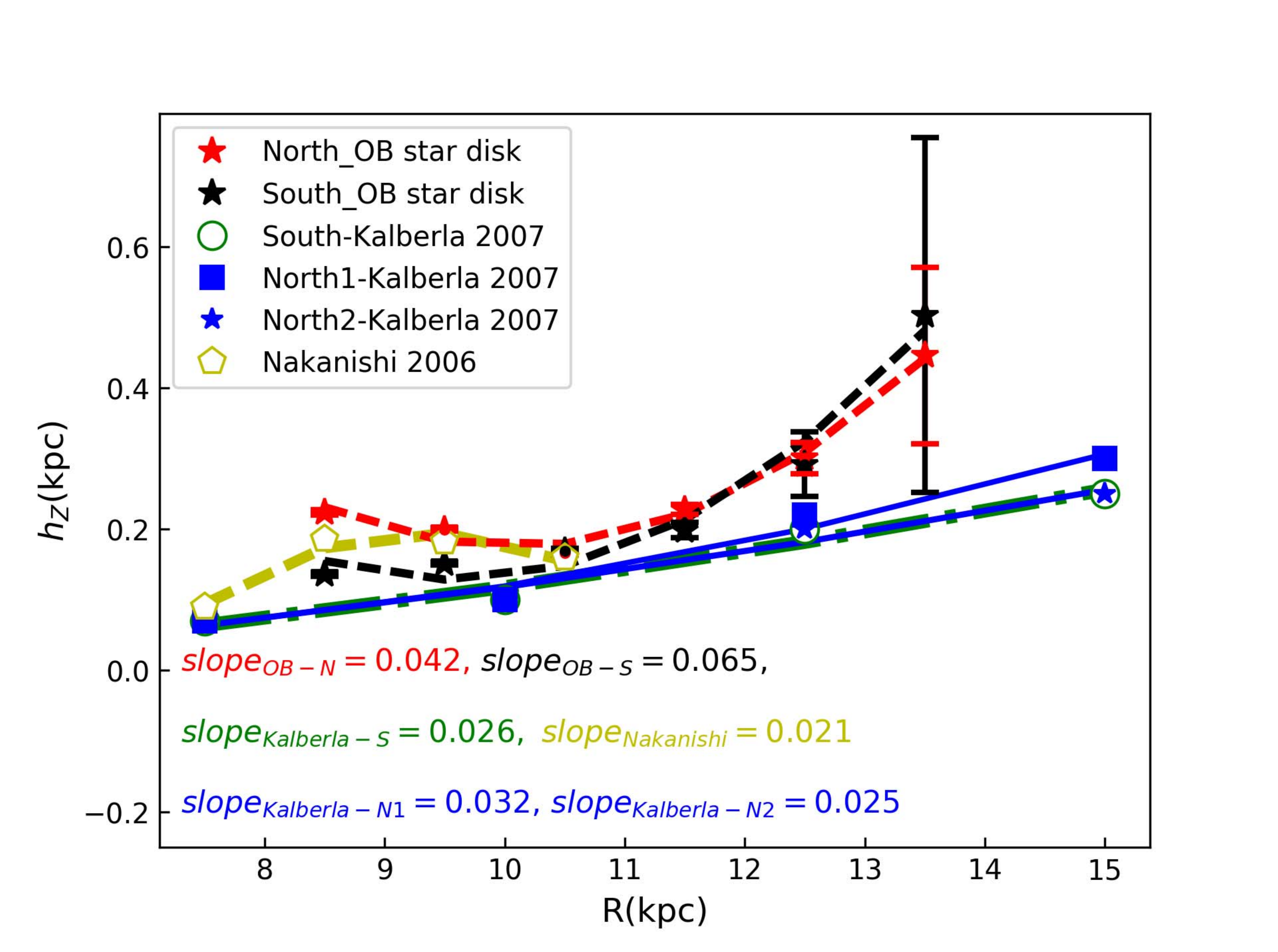}
  \caption{The figure shows the OB-type stars scale height distribution, compared with the gas disk, along the radial distance, red line is north side scale height, black line is the southern scale height. Green line is the \citet{Kalberla2007} southern scale height of the HI gas disk, the blue rectangles and pentagram are also the \citet{Kalberla2007} northern HI thickness distribution, yellow is corresponding to the H$_{2}$ results in \citet{Nakanishi2006}. {\bf The average slope of the scale height (h$_{z}$) vs. distance (R) }describing the strength of flaring is labeled in the bottom left.}
  \label{flaringns}
\end{figure}

The scale height of the young stellar disk compared with HI and H$_{2}$ gas disk of \citet{Kalberla2007, Nakanishi2006} is shown in Fig.~\ref{flaringns}, for \citet{Nakanishi2006} the data is mainly covered by $b$ = [-1.5 1.5$^\circ$] and $l$ = [0 to 90 $^\circ$, 270 to 360 $^\circ$] and for \citet{Kalberla2007} the data range is covered by all sky disk region by discarding $b$ $>$ 30$^\circ$ and Galactic distance $R$ = [-3.5 3.5 \,kpc]. It is shown that the young stellar disk, for the overall trend, is slightly thicker than the gas disk but it is not significantly different for the scale height, which is expected since that we declare that OB-type stars could be used to traced more gas properties and dynamics, these will be shown in our series of works. The OB-type stars are very young and have not moved out of the star formation area so that there is no doubt the structure of the OB-type stars are inherited some properties of the gas clouds that are forming stars. We also find that the HI disk has some differences with the H$_{2}$  disk due to the temperature difference possibly. The distance of our OB star sample is from Gaia, the extinction is very small and we could use its precise distance within 8-14 \,kpc. If we agree there are some influence on our results caused by extinction and not perfect selection effects, then the error bar will be enlarged, thus then the true difference will possibly be reduced so that strengthen our points to some extent, that is, we could use OB star to infer some properties of the gas disk, then it will encourage us to work more.

The most important tracer for molecular gas in galaxies is the CO spin line of different isotopes, especially the 2.6 \,mm line of CO and high-sensitivity CO observations are essential for describing the structure of the disk in galaxies, sometimes we also choose to convert CO to H$_{2}$. In \citet{Sun2019}, they also confirmed the gas thin disk (FHWM $\approx$ 90 \,pc) and revealed the gas thick disk was around 280 \,pc consistent with the HI gas disk scale height (250$-$300 \,pc), meanwhile they pointed out molecular gas properties might be related to massive stars evolution, which support our points that the gas disk properties could be inherited by the young OB-type stars so that the young stellar disk could be compared with gas disk to investigate more intriguing details. Recently, \citet{Su2021} find that the molecular gas disk has two components, gas thin disk and thick disk, with scale height of 85 \,pc and 280 \,pc in the region of $l$ from 16$^\circ$ to 52$^\circ$ and $b$ from -5.1 to 5.1$^\circ$, respectively, these values are belonging to our range of OB stellar disk. We will have a few more discussions in the latter part.

\subsection{The scale length of the OB-type stellar disk compared with gas disk}\label{subsec:figures}

\begin{figure}
  \centering
  \includegraphics[width=0.5\textwidth]{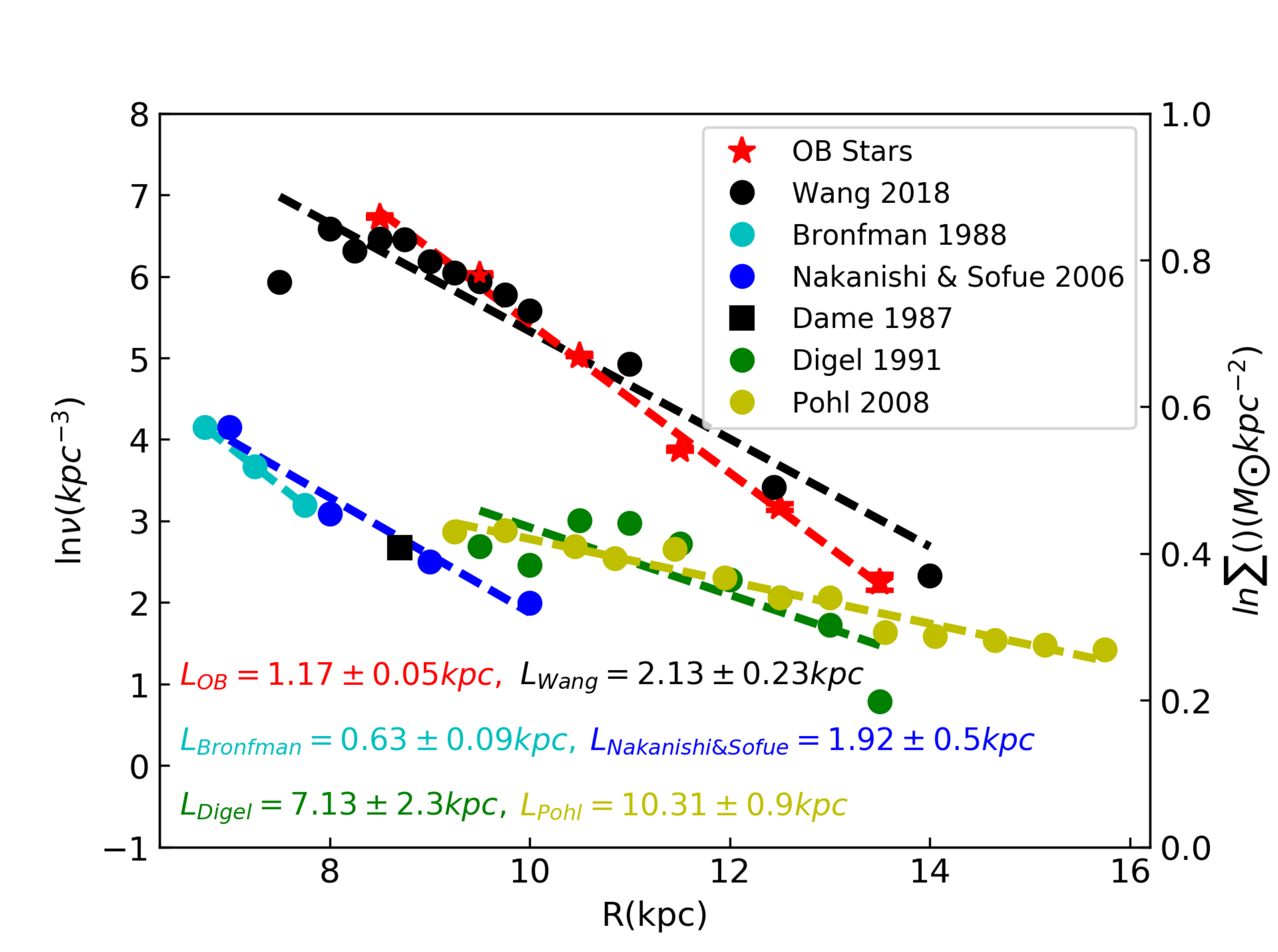}
  \caption{The figure shows mid-plane density distribution along the distance, the linear fitting results are shown as dashed lines with different colors, the scale length of different works for H$_{2}$ gas disk \citep{Dame1987, Bronfman1988, Digel1991, Nakanishi2006, Pohl 2008} is labeled in the bottom left, the H$_{2}$ value is converted from CO \citep{Heyer 2015}. The left y axis is corresponding to the stellar disk and the right axis is the gas surface density. The gas density is adjusted by shifting up 5.5 and the \citet{wang2018b} is shifted up 10 in order to compare in one figure.}
  \label{scalelength}
\end{figure}

The stellar density of the disk, general speaking, is exponentially distributed in the radial direction and gradually decreases with the increase of the distance and we fit the logarithm of the stellar density in the direction of the Anticenter belonging to the radial range of R = 8$-$14 \,kpc. The scale length of the OB-type star is 1.17 $\pm$ 0.05 \,kpc and the comparison with previous works of the H$_{2}$  gas disk which has been transferred from CO data \citep{Dame1987, Bronfman1988, Digel1991, Nakanishi2006, Pohl 2008} is shown in Fig.~\ref{scalelength}, it should be pointed out that, for gas density, we directly use the data points from other literatures. Some works have number density with units cm$^{-3}$ and this gas density is often calculated by column density divided by radius size, however, here we have actually used the surface mass density (M$_{\sun}$pc$^{-2}$ )which is also suitable for us to compare scale length. In order to compare with stellar disk in one figure, the gas density value and stellar density value of \citet{wang2018b} is shifted up in y-axis. It shows that the young stellar disk scale length is shorter than the gas disk when comparing the value labeled in the figure except the work of \citet{Bronfman1988} with only three points in a narrower range {\bf so it is not reasonable, others are implying that generally the gas disk is more extended than the stellar disk}. We could also attempt to deduce from here stellar disk might be more compact, similar to the hot dust, than the gas and cold dust disk. For the gas disk region mentioned in this part, the range of \citet{Dame1987} is covered by 10 $-$ 20$^\circ$ in latitude at all longitudes and all or nearly all large, nearby clouds at higher latitude. Latitude is from -2 to 2$^\circ$ and longitude is from 300 to 348$^\circ$ is corresponding to the region of the \citet{Bronfman1988}, but we only use the north data beyond 8 \,kpc. Based on the smoothed$-$particle hydrodynamical simulations and the complied data \citep{Dame2001}, \citet{Pohl 2008} made full use of the CO survey of which total area covered by the composite survey is 9,353 deg$^2$ $-$ more than one$-$fifth of the entire sky and nearly one half of the area within 30$^\circ$ of the Galactic plane. \citet{Digel1991} used data located in the outer galaxy within the disk region from $l$ =[65 116$^\circ$]. Here H$_{2}$ gas surface density is converted from CO surveys, using uniform assumptions regarding the Galactic rotation curve, solar radius, and the CO$-$to$-$H$_{2}$ conversion factor, all these data are from the review \citep{Heyer 2015}.

In addition to that, we notice that the scale length of the OB-type stellar disk here is different from the results given by \citet{Lichengdong2019}, it is $2.10 \pm 0.01$ \,kpc and scale {\bf height is from 132$-$450 \,pc, both are morphological or geometrical definition, we suggest it is} due to that  \citet{Lichengdong2019} used Gaia DR2 and 2MASS photometric data and it must include the early type and late-type of OB-type stars, what we want to emphasize here, our sample is lacking of late type ones and mainly consisted of early type stars caused by the selection methods, so the scale length has some differences for these two works. 

\subsection{The mid-plane displacements of the disk}\label{subsec:figures}

The gas warp was detected as shown in  \citet{Kerr1957, Bosma1981, Briggs1990}. Recently, with the help of classical Cepheids, \citet{Chen2019} have revealed an intuitive three-dimensional (3D) map of the stellar warp by star counts methods, it is covering the all sky until 20 \,kpc. Meanwhile,  \citet{Skowron2019a, Skowron2019b} have also showed us number density stellar warp signals in different distance and azimuthal angles located in the north and south sky and declared that the amplitude of its northern part is very prominent and stronger than that of the southern part. Meanwhile, the kinematical and height signal of the stellar warp and gas infall origins are also unraveled in \citet{wang2020d} with the help of sample mainly from the LAMOST Galactic-Anti Center in the north hemisphere. Some other works about warp signals with different tracers and methods could also be found in \citet{Poggio2018} by using Gaia sample mainly in the north side, then \citet{Poggio2020} used all sky Gaia 12 million giant stars to detect the warp precession for the first time. However, \citet{Zofia2021} recently recalculated the warp precession and found that there is no need for precession using different warp parameters but same approach and Gaia DR2 kinematic data.

\begin{figure}
  \centering
  \includegraphics[width=0.5\textwidth]{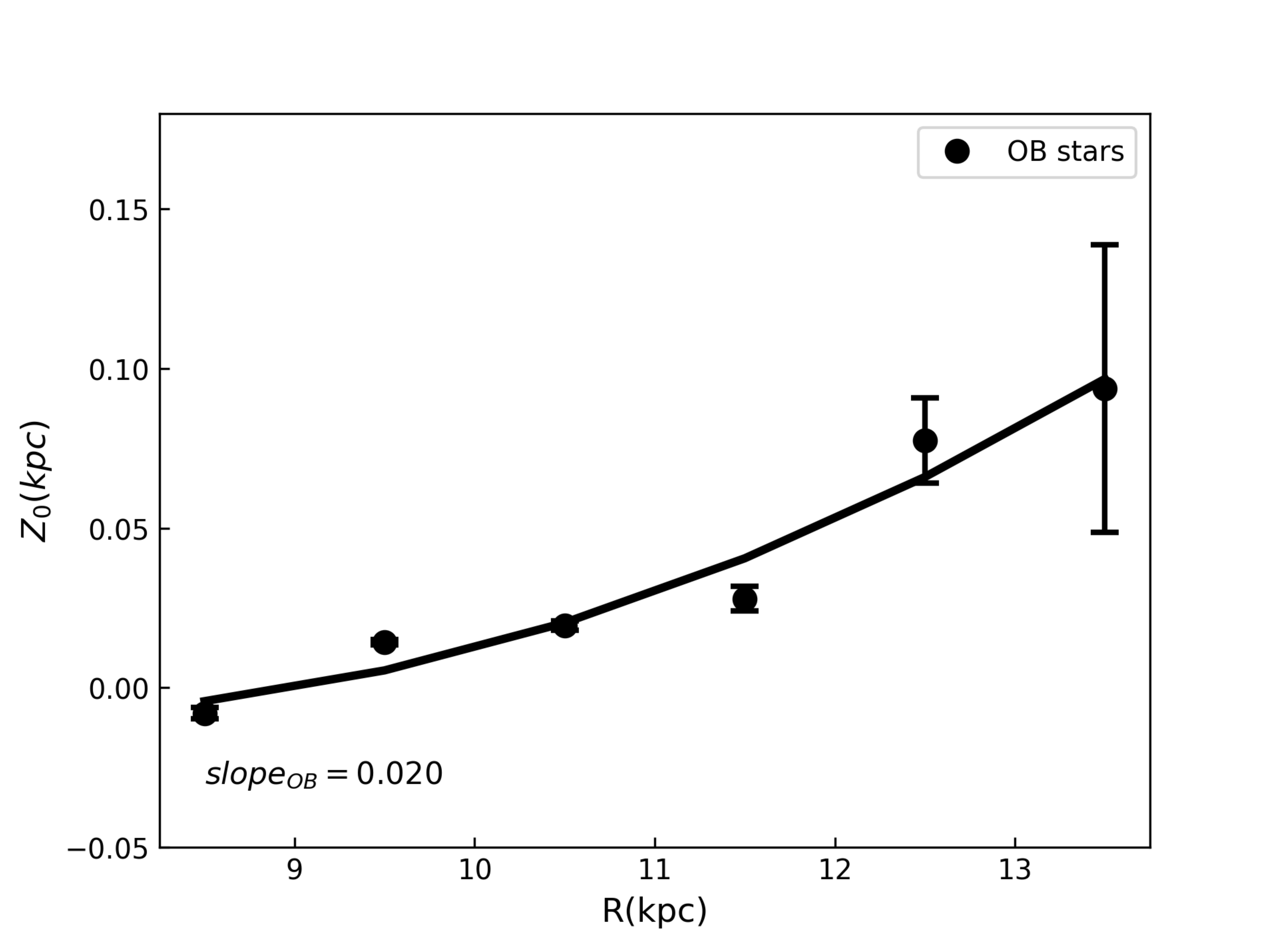}
  \caption{The figure displays the mid-plane displacement distribution along the $R$, almost are within the 100 \,pc and the slope shown in the bottom left is 0.020.{\bf please notice that the slope is just for quantitative analysis, which is different from flaring strength.}}
  \label{z0R}
\end{figure}

During this work we introduce the mid$-$plane displacement ($Z_{0}$) to Equation \ref {numbermodel} and Equation \ref {lnlike} to do the decomposition and fit the OB-type stars in radial slice displayed in Fig.~\ref{z0R}. Intriguingly, we find that the mid-plane displacements are existing and different locations have different $Z_{0}$ values, but almost are within 100 \,pc accompanying by that the value is approximate to be 0 around solar location in the range of 8-9 \,kpc. 

More importantly , we find the offset is increasing with the distance and infer that there is a possibility it might reflect suspected signals like the warp here. If it is true we could say this work detects the similar warp signal in the young stellar population and warp should happen in the early time. However, here from only one figure, we could not rule out  other possibilities due to the limited range of the sample and other possible perturbations. In short, the pattern might be caused by warp, or caused by the heating of some perturbations which are not the same with warp dynamical mechanisms, sometimes we actually could not discriminate whether the vertical signal is from warp or external perturbations \citep{wang2020a, wang2020c} or other unknowns, which is also not the main target of this paper. Please notice that the zero point bias might affect the exact value here but it will not change the pattern shown in this work. The significance of the signal is still needed to explore more in the future, but we just don't want to rule out some possibilities here, our motivation is flaring.

By using the similar method introduced in \citet{Xu15}, {\bf \citet{wang2018b} proposed that some disk oscillations are very complicated but could still be explained, to some extent, by shifting either the thin or the thick disc, so during this work we do like this in the young stellar disk is reasonable and the value of the mid$-$plane displacement ($Z_{0}$) is also not strange at all.} In general, our mid-plane shifting pattern is actually similar with that in Fig. 5 of recent work finished by \citet{xu2020} based on the kinematics method.

\subsection{Discussions}\label{subsec:figures}

\subsubsection{ Comparisons for the disk w/o North and South sides}\label{subsec:figures}

\begin{figure}[htbp]
\centering
\subfigure{
\includegraphics[scale=0.127]{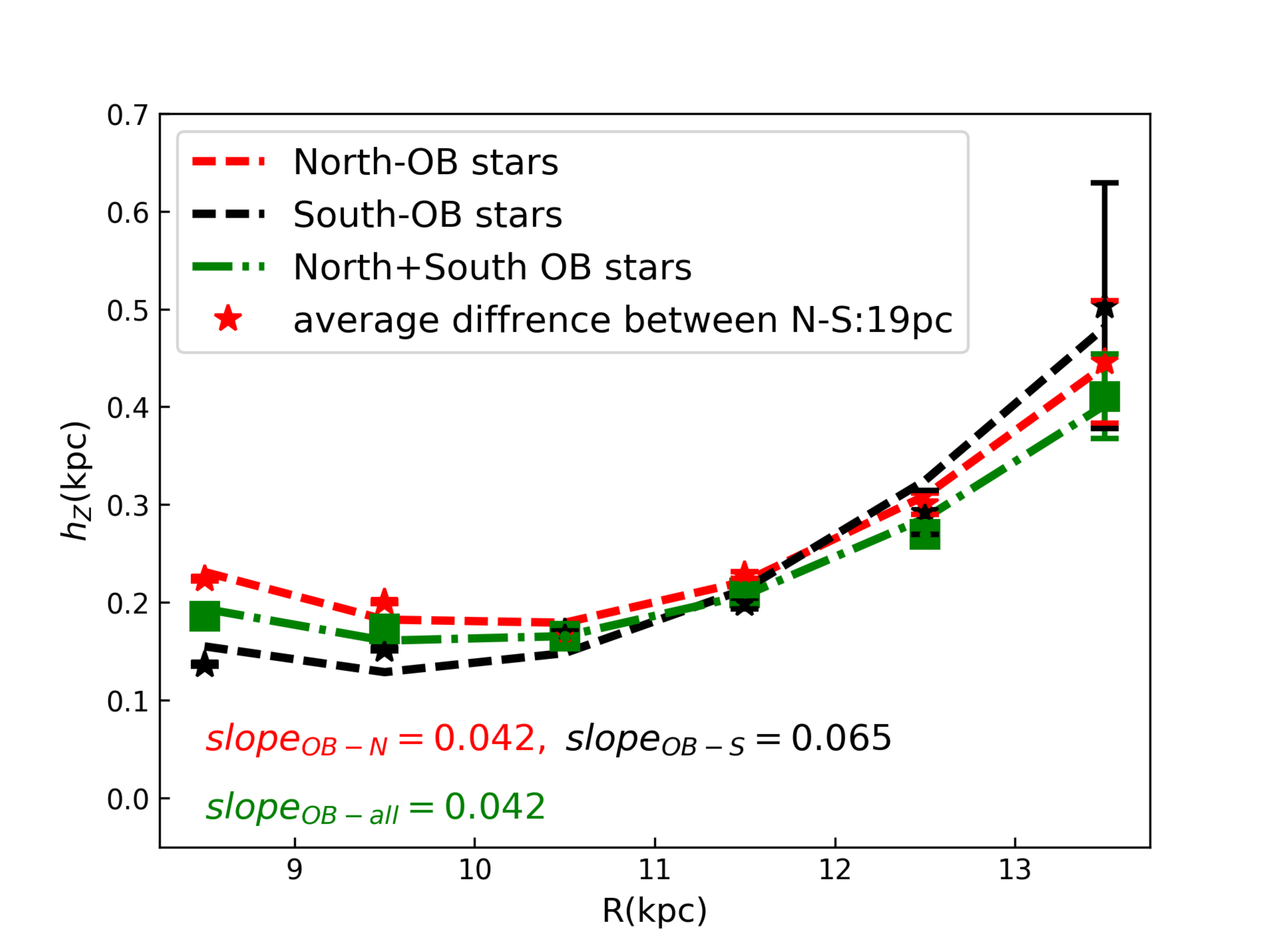} \label{1}}
\quad
\subfigure{\includegraphics[scale=0.13]{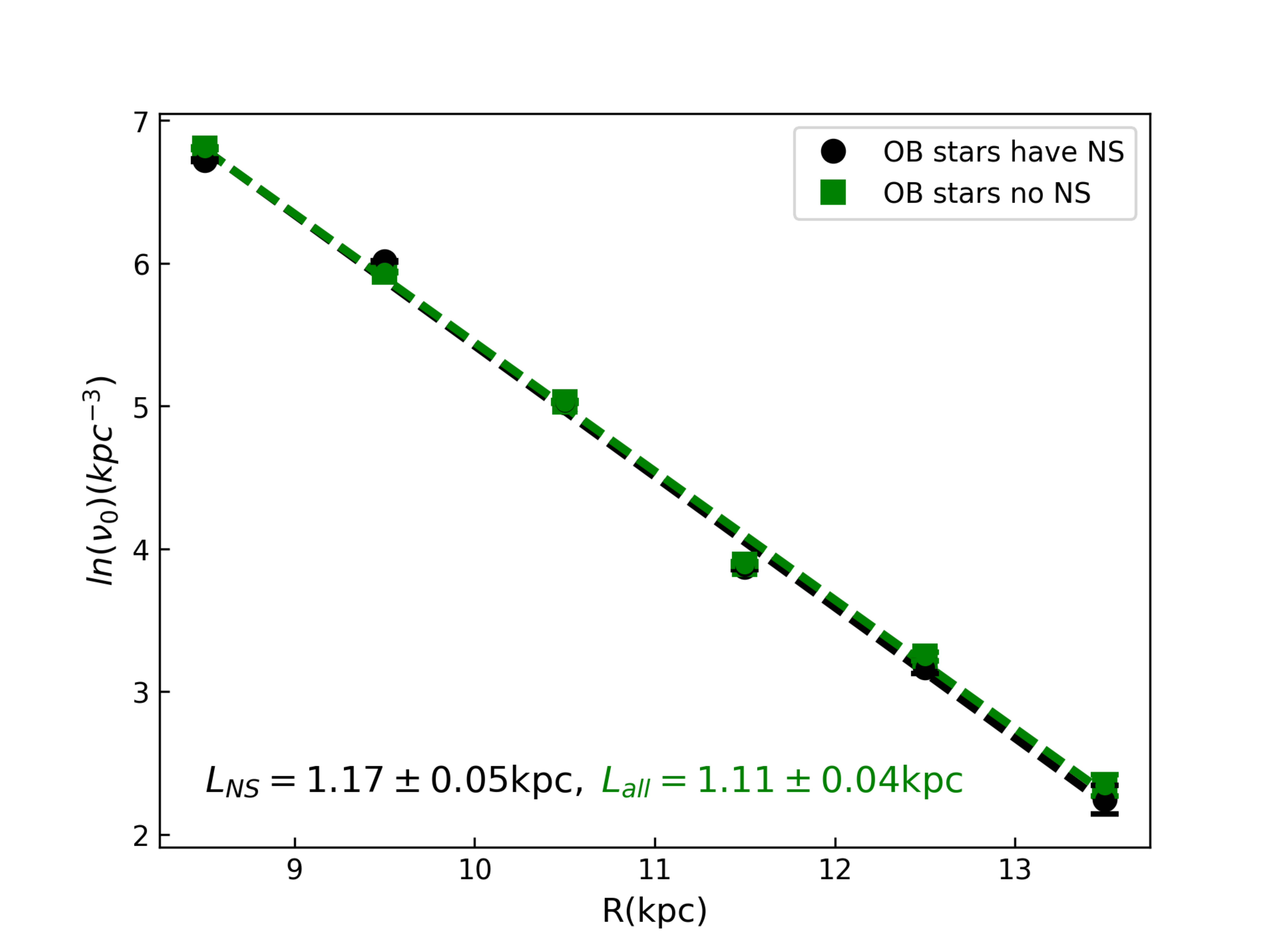} \label{2} }
\quad
\subfigure{\includegraphics[scale=0.126]{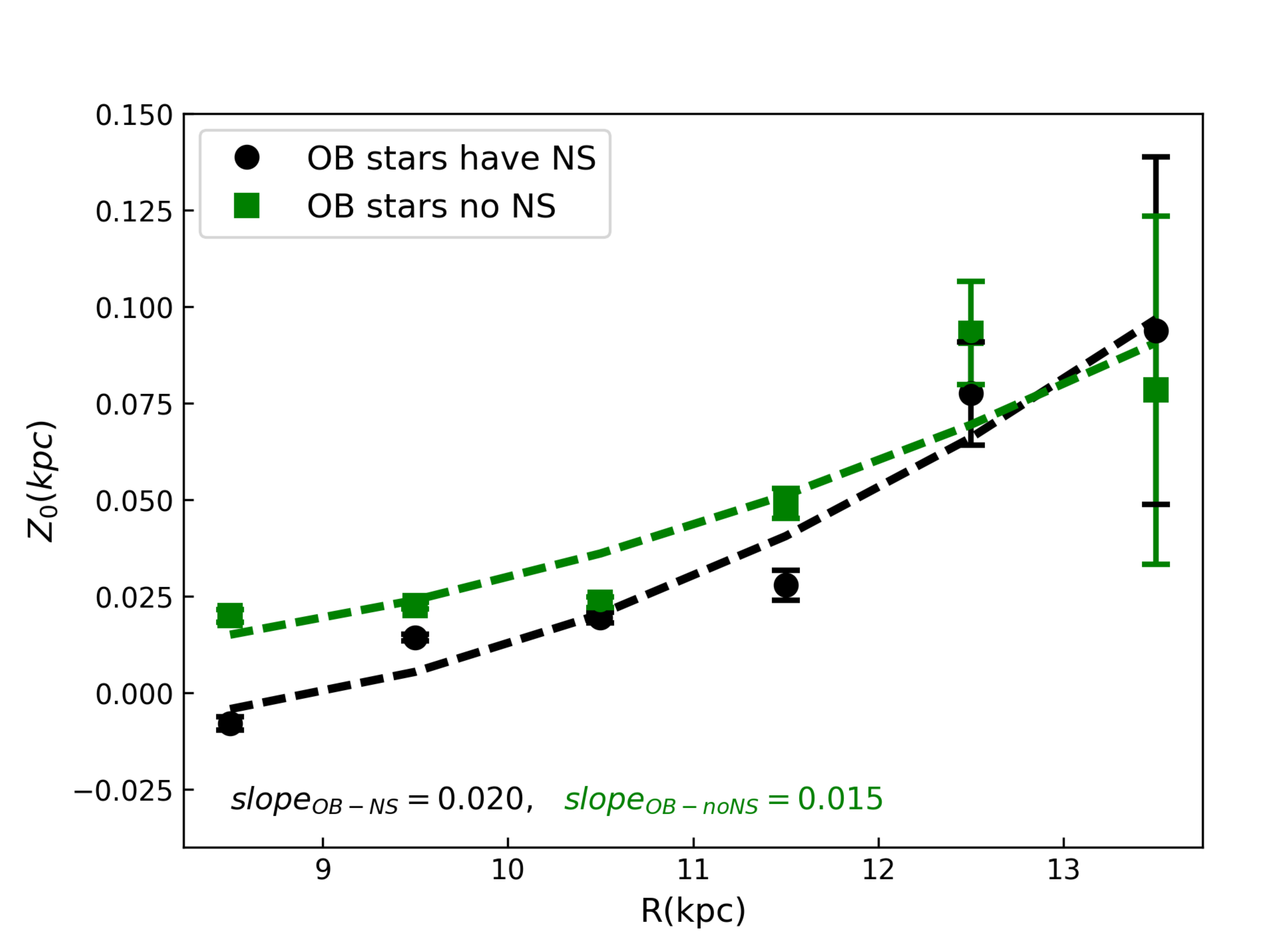} \label{3} }
\caption{Comparison for the disk in consideration of north and south sides and the one without considering the scale heights of north and south sides simultaneously, which are colored by red or black pentagram and green square respectively. The top one is the scale height, the middle one is the mid-plane density and, the slope of the scale height and scale length {\bf along the distance} is labeled. The bottom one is the mid-plane offsets with its slope label, clearly, the pattern is stable.}
\label{NSno}
\end{figure}

In order to test the effect of the method with decomposition of the north and south sides, we have finished a comparison for the disk features between the north-south scale heights and the values without considering north-south sides. As shown in Fig.~\ref{NSno}, the scale heights, scale lengths and mid-plane offsets are compared from the top to the bottom. For the {\bf scale heights, there are no large differences and the general pattern} is the same with each other and similarly, the mid-plane offsets are also showing the same trend with some differences. Interestingly, the mid-plane density and the scale length are matched very well. All these show that our method with north and south sides will not change our main conclusion for this work, especially the flaring.

\begin{figure}[htbp]
\centering
\subfigure{
\includegraphics[scale=0.4]{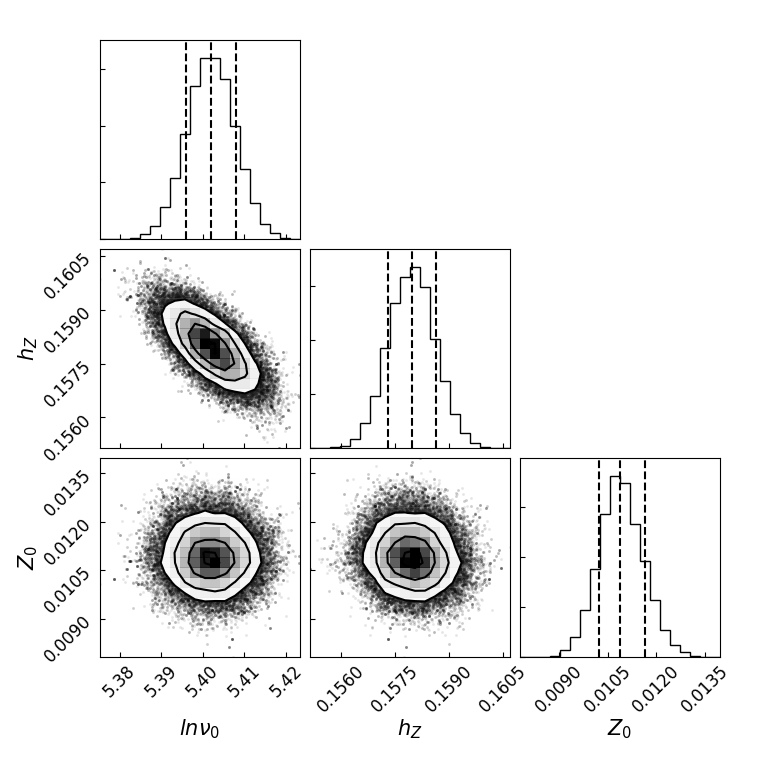} \label{1}}
\caption{The panel shows the likelihood distribution of the parameters ($ln (\nu_{0})$ ($kpc^{-3}$),  $h_{z}$ (\,kpc), $Z_{0}$ (\,kpc)) for 8$-$14 \,kpc data drawn from the MCMC simulation without considering north and south sides simultaneously, the uncertainties in the estimates are determined using the 15th and 85th percentiles of the MCMC samples.}
\label{mcmcnoNS}
\end{figure}

By combining all stars belonging to 8-14 \,kpc without radial slices, we have also finished the fitting results for the young stellar disk, the scale height of the young stellar disk is 0.16 \,kpc and the mid-plane offset of the disk is 11 \,pc, the fitting results are quite well and shown by us in Fig.~\ref{mcmcnoNS}.

\subsubsection{Possible scenarios for the flaring }\label{subsec:figures}

With the release of Gaia DR2 data, \citet{Antoja2018} found phase mixing features of the Milky Way disk possibly due to the disturbance caused by the vibration of the dwarf galaxy passing through the disk 300-900 \,Myr ago. The lifetime of OB stars is mostly $<100$ \,Myr and the gas disk forming the OB-type star could possibly be disturbed by the dwarf galaxy so that some imprints, might be inherited by the OB-type stars \citep{Cheng2019}. \citet{Cheng2019} also introduced two types of interference sources: First, internal disturbances, such as spiral structures, huge molecular clouds, or central rods; The second, is external disturbances, such as dwarf galaxy or satellites passing through the Milky Way. In addition to the aforementioned disturbances, \citet{Cheng2019} proposed another case, that is saying, a few decades ago, the dark matter sub-halo perturbed the disk and left marks on these young stars. Similar perturbation scenarios to the disk such as warp, stream, satellites, dwarf galaxies, dark matter sub-halos are also mentioned in \citet{wang2018a, wang2019, wang2020a, wang2020d, Khanna2019, BlandHawthorn2019, Lopez2020, xu2020} and reference therein. 

{\bf Flaring is possibly related to the disk heating, there are many mechanisms proposed in the past many years}. e.g. Spiral arms could be the important contributor to the in-plane motions but without influence on the vertical motions \citep{Sellwood1984, Sellwood2013}. Giant Molecular Clouds (GMC) could heat the disk and redirect the velocity of stars out of the plane \citep{Lacey1984}. Bar might not be important beyond solar neighborhood but can not be ignorable in the inner bulge region \citep{Moetazedian2016}. Migration might heat the disk and contribute to the flaring, which was mentioned in recent work \citep{bovy2016}, but the different scenarios proposed by \citet{Minchev2012, Minchev2018} suggest that migration is not important for heating and flaring. External perturbations such like dwarf galaxy was heating the disk \citep{House2011} and is the mechanism for the flaring \citep{Kazantzidis2008}, in addition, mergers and perturbations from satellite galaxies and subhalos can induce radial mixing in the outer disk \citep{Quinn2009}. {\bf For our work, we focus on the average slope of scale height vs. distance which is used to describe the strength of the flaring features in different populations}, then we could attempt to infer which mechanism of the flaring is more important, we agree the different heating mechanisms have different size of the thickness but it is not our target of the paper, hopefully in the future we could combine kinematics and star counts to pursue more details about this intriguing question.

Here again, as we mentioned in the introduction, we attempt to classify the mechanisms of the flaring into two types: one is the secular evolution of the disk, which expects the younger population has weaker flaring strength than the older population, the other one we think it is perturbation caused by many agents, which don't expect the population evolution effect with age for the origins of the flaring. Against this big background then we could discuss which one is more important, {\bf we also know the physical details might be more complicated and we are not planing to rule out other physical scenarios only from current analysis}.

Therefore, from current results, we clearly detect the flaring strength of young OB stellar population is similar with or even slightly stronger than the old RGB population by focusing on the thin disk by using the slope of the scale height vs. distance to describe the strength of the flaring, with the average value of the slop is 0.042 (north-OB) vs. 0.065 (south-OB) vs. 0.020 (RGB),  the qualitative conclusion of which is similar with \citet{wang2018b}, they also found the flaring strength of  Red Clump Giant stars (RCG) is similar with the RGB pattern for the thin disk. By comparing flaring strength of the thin disk and thick disk, both \citet{wan2017} and \citet{wang2018b} detected the similar flaring features for these two different populations. {\bf That is to say, we suggest that the secular evolution is not the main contributor during our works and for the quantitative analysis of \citet{wang2018b} shown in this paper, it is the same value of the slope for the thin and thick disk, as shown in Fig.~\ref{flaringns1}.}

And for the current work, if we look carefully the Fig.~\ref{flaringns1} by only focusing on the thin disk stellar population, the purple solid line is the F/G type stars, the green line with triangles is K-type/RGB stars, the red and black line with pentagram is OB-type stars, it is not possible for us to see the trend that the older the populations, the stronger the flaring by focusing on the variation of the scale height or {\bf the slope of the scale height with distance}, the quantitative analysis for the strength of the flaring is that the OB type is 0.042 or 0.065, F/G type is 0.145, K type is 0.020, which is also not consistent with the scenario of the secular evolution.  Meanwhile, if we concentrate on the thick disk stellar population, the blue line is F/G type stars, the green line with rectangles is RGB/K type stars, the former one is generally younger than the latter one, but it appears that both have similar flaring features with the slope is 0.023 and 0.020 separately, it is hard to be explained by secular evolution scenarios.  All these evidence again support that the secular evolution predicting the flaring strength is different in different populations with that older stellar population should have stronger flaring, is not the main contributor, although we could not completely rule out the contributions of this scenario up to now. In other words, the perturbations such as dwarf galaxy might be more possible according to current two classifications.

Recently, based on the smooth particle hydrodynamical simulations with clumpy episodes, \citet{Silva2020} discovered broken exponentials is existing in the low-[$\alpha$/Fe] population and more importantly, flare significantly for low-[$\alpha$/Fe] stars but only weakly for high-[$\alpha$/Fe] populations. Meanwhile they also found that for low-[$\alpha$/Fe] thin disk stars, the flaring level decreases with age, which is also contrary to the secular evolution and it is similar with GMC perturbation scenario.

With the help LAMOST K giant stars and two types of simulations, \citet{xu2020} pointed out the external perturbations might influence the outer disk beyond solar neighborhood and revealed that the flaring could be explained by perturbations qualitatively, which supports our viewpoints here about the flaring. Combining young OB type Stars and old K giant stars we {\bf could infer that the sensitive time of the flaring features to the possible perturbations could be from very early time in OB population to the relatively late time in K giant population.}

Although we are far from the exact flaring mechanisms, we suggest here, possible perturbations such like the mentioned above could heat the disk and then, cause the disk flaring, therefore we could detect that the disk displays clear flaring signatures. More importantly, during this work we discover these possible heating mechanism might be thickening the disk of both sides in some similar ways, especially in the outer disk,  this will push us to have better knowledge of the well$-$known flaring disk, especially for theorists.  

\subsubsection{ Relations between the young stellar disk and gas disk}\label{subsec:figures}

We also find the evidence of relationship between the young OB-type stars and the gas disk by comparing their structure parameters. It is shown in Fig.~\ref{flaringns} that the scale height of the OB-type stars is slightly thicker than the gas thickness by 9 pc in the south and 33 pc in the north for the median value, the small difference supports the truth that OB stars is very young and only last a few million years and during this time the stars should not have had time to significantly move from their birth locations, so the young stars and gas could share similar properties and we could use young OB stars to trace some physics about gas. For the reason that the OB stellar disk is a little thicker than the gas disk, we speculate there are two reasons, one is that the stars are hotter than gas so it is thicker, the other is the stars are perturbed by some agents and they are possibly more effective for stars, the latter one is also consistent with the perturbations scenarios if we assume that the OB stellar disk is heated by perturbations after it formed from the gas. As far as we know, we don't see some other works show the similarities of the young stellar disk and gas disk so that it might encourage our community to explore more about these. {\bf We are planing to investigate more and farther for stellar, gas, and dust disk. For this work, due to the sampling rate is low when the distance is larger than 14 \,kpc so we only focus a limited range here.}

In short summary, we detect the clear flaring signatures of the young stellar disk that is possibly mainly contributed by perturbations, this symmetrical flaring evidence for young stellar disk is clearly of vital importance for us to understand flaring mechanisms and even gas disk physics.

\section{Summary} \label{sec:highlight}

In this work, we explore the spatial structures of young stellar disk with the LAMOST DR5 OB-type stars catalog, based on the Bayes method for selection effect correction and non-parametric method for disk decomposition, {\bf both are robustly tested in our previous works}.  We discover that the north-south flaring in the outer disk is symmetrical and the scale height is {\bf from 0.14 to 0.5 \,kpc at different distance from 8 to 14 \,kpc}. The flaring strength of the younger stellar disk is similar to or slightly stronger than the results of the thin disk traced by older red giant branch stars {\bf with the average slope charactering the strength is 0.042 or 0.065 (OB) and 0.020 (RGB) separately}, which is supporting that the flaring might not be mainly contributed by the secular evolution scenarios, which are predicting the flaring of the old populations should be stronger than the younger ones. And meanwhile, we speculate that the mechanism might have similar influence on the both disk sides in some similar ways.

{\bf We also find that the OB stellar disk is slightly thicker than the gas disk by 9 \,pc in the south and 33 \,pc in the north for the median value,} implying that we could attempt to, to some extent, use OB-type stars to infer the gas properties in the future and what's more, we unravel that the stellar disk might be more compact than the gas disk by showing that the scale length of OB-type stars is shorter than that of gas. We believe that more comparisons of the young stellar disk and gas disk will be non-trivial to uncover more secrets of the formation of the disk.

{\bf Finally,  it is displayed that the mid-plane displacements($Z_{0}$) of the thin disk have a clear increasing trend with distance from 8-14 \,kpc, that is possibly implying the signal like warp, but so far we are still not sure its origins and significance due to that we can not discriminate the warp contributions and perturbations contributions, or other unknowns from the current limited range and analysis. All these observational evidences of the young stellar thin disk shown here will encourage us to reveal more about the flaring mechanisms and gas disk physics, more works will be shown in the future.}

 \acknowledgements
We would like to thank the anonymous referee, Lixia Yuan, Yang Su, Miaomiao Zhang, Ningyu Tang, Yan Sun, Ivan Minchev, Leandro Beraldo e Silva for his/her very helpful and insightful comments and discussions.This work is supported by the National Key Basic R\&D Program of China via 2019YFA0405500, 2019YFA0405501, the National Natural Science Foundation of China (NSFC) under grant 11773009, 11873057, 11390371, 11673007, 12003027. W.Y.C. is also supported by the "333 talents project" of Hebei Province. H.F.W. is also supported by the LAMOST Fellow project, funded by China Postdoctoral Science Foundation via grant 2019M653504 and 2020T130563, Yunnan province postdoctoral Directed culture Foundation, and the Cultivation Project for LAMOST Scientific Payoff and Research Achievement of CAMS-CAS. The author gratefully acknowledges the support of K.C. Wong Education Foundation. The Guo Shou Jing Telescope (the Large Sky Area Multi-Object Fiber Spectroscopic Telescope, LAMOST) is a National Major Scientific Project built by the Chinese Academy of Sciences. Funding for the project has been provided by the National Development and Reform Commission. LAMOST is operated and managed by National Astronomical Observatories, Chinese Academy of Sciences. This work has also made use of data from the European Space Agency (ESA) mission {\it Gaia} (\url{https://www.cosmos.esa.int/gaia}), processed by the {\it Gaia} Data Processing and Analysis Consortium (DPAC, \url{https://www.cosmos.esa.int/web/gaia/dpac/consortium}). Funding for the DPAC has been provided by national institutions, in particular the institutions participating in the {\it Gaia} Multilateral Agreement.

\end{CJK*}
\end{document}